\documentclass[twocolumn,showpacs]{revtex4}
\usepackage{graphicx}
\usepackage{dcolumn}
\usepackage{bm}
\usepackage{amsmath}
\usepackage{amsfonts}
\usepackage{amssymb}

\providecommand{\U}[1]{\protect\rule{.1in}{.1in}}

\begin{document}

\title{Polytropic process and tropical Cyclones}
\author{Alejandro Romanelli}
\altaffiliation{alejo@fing.edu.uy}
\author{Italo Bove}
\author{Juan Rodr\'{\i}guez}
\affiliation{Instituto de F\'{\i}sica, Facultad de Ingenier\'{\i}a\\
Universidad de la Rep\'ublica\\
C.C. 30, C.P. 11300, Montevideo, Uruguay}
\date{\today }

\begin{abstract}
We show a parallelism between the expansion and compression of the atmosphere in
the secondary cycle of a tropical cyclone with the fast expansion and compression of wet air in a bottle. We present a simple
model in order to understand how the system (cyclone) draws energy from the air humidity. In particular we suggest that the upward (downward) expansion (compression) of the warm (cold) moist (dry) air follows a polytropic process, $PV^\beta$= constant. We show both experimentally and analytically that $\beta$ depends on
the initial vapor pressure in the air. We propose that the adiabatic stages in the Carnot-cycle model for the tropical cyclone be replaced by two polytropic stages. These polytropic processes can explain how the wind wins energy and how the rain and the dry bands are produced inside the storm.
\end{abstract}

\maketitle

\section{Introduction}

Observations and numerical simulations of tropical cyclones expose the importance of latent heat flux from the sea
surface to understand the development and maintenance of these intense storms. The self-induced heat transfer from the
ocean can be understood through a simple thermodynamical model \cite{Emanuel}, that uses the evaporation of water from the ocean surface as the primary
source of power of the tropical cyclone.

The three-dimensional wind field in a tropical cyclone can be characterized by a primary and a secondary
circulations. The primary circulation is the rotational part of the flow, it is essentially a purely circular movement in the horizontal plane. The secondary
circulation is in the radial and vertical directions. The primary circulation has the strongest winds and is responsible
for most of the damage that the storm causes; while the secondary circulation is relatively slower, but it rules the energy distribution in the storm.
In the last 30 years, the energy dynamics of the secondary circulation has been modeled by an atmospheric Carnot cycle.  This cycle of the air flow together with the empirical knowledge of the dependence between the heat delivered by the sea and the speed of the winds,
provides a fundamental result for the upper bound of the wind speed that a storm can attain \cite{Emanuel}. 

On the other hand, we recently studied both theoretically and experimentally the thermodynamics of the fast expansion
($\sim0.1$ s) of wet air in a bottle \cite{romanelli}, taking into account the expansion
of wet air, vapor condensation, and the corresponding latent heat. We set up a simple experimental
device with a bottle containing compressed air and liquid water and verified that the gas expansion in the bottle is well approximated by a polytropic
process $PV^\beta=$ constant, where the parameter $\beta$ depends on the initial conditions. We also find an analytical
expression for $\beta$ that depends only on the initial vapor pressure in the air (thermodynamic initial conditions) and
is in good agreement with the experimental results. In the present paper we show a parallelism between the expansion and
compression of the atmosphere in the secondary cycle of the tropical cyclone with the fast expansion and compression
of air in a bottle. We propose to substitute the adiabatic processes in the Carnot-cycle model of the tropical cyclone
by two polytropic process of expansion and compression of wet air. These polytropic processes can explain the energy gain by the wind
and the appearance of rain and cold dry bands in the tropical cyclone.

The paper is organized as follows, in the next section we present our experiment. In the third section we
develop some topics about the polytropic process and present a theoretical expression for the polytropic exponent for wet air. In the fourth section we present the maximum potential intensity in the modified Carnot engine. In the last
section some conclusions are drafted.

\section{Experimental device}

What primarily distinguishes tropical cyclones from other meteorological phenomena is the condensation as a driving
force. A fast expansion of air with water vapor produces vapor condensation with the corresponding liberation of latent heat. 

In this section we present an experimental device where the movement of a fluid (which  responds to hydrodynamic
equations) concordantly with the rapid expansion of moist air, produces water vapor condensation in the air (a thermodynamic process).
We find here the essential elements of the moist air expansion in the tropical cyclone, that is a toy model for this storm.

\begin{figure}[th]
\begin{center}
\includegraphics[scale=0.8, angle=0]{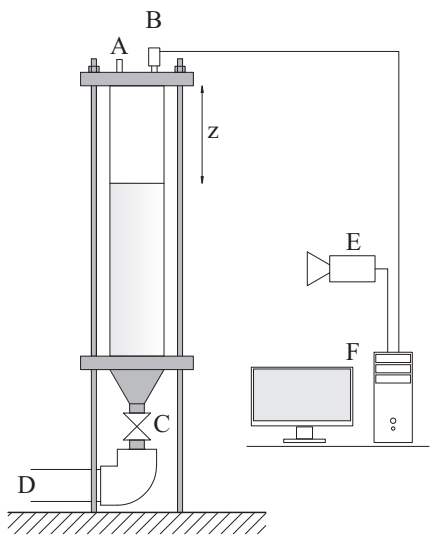}
\end{center}
\caption{The stationary bottle, (A) is a tire valve, (B) is a pressure manometer, (C) is a faucet, (D) is a pipe to the
drain, (E) is a film camera, (F) is a computer with a data acquisition card and Z is the height of the gas column
situated over the water column.}\label{f1}
\end{figure}

The experimental set up consists in  a transparent
acrylic cylinder of inner diameter $5.40$ cm, wall thickness $ 3.0$ mm and length $1.0$ m, that is maintained in vertical
position solidary to the laboratory as depicted in Fig.~\ref{f1}. The upper and lower covers are both of aluminum. The
upper one has a tire valve and a digital manometer (Nuova Fima TP ST-18) working from $0$ to $10$ bar. The lower cover
has a conical shape with a slope of $30^{\circ}$ with a $3/4$" faucet that allows to fill and evacuate the water in the
bottle. In a typical trial the bottle is filled with about $1400$ cm$^3$ of water and the rest with air. The compressed
air is introduced through the valve. The inner temperature is monitored with a thermistor. We wait some minutes until the
system reaches thermal equilibrium and then the faucet is opened. The level of the gas is filmed with a camera Photon
Focus DS1-D1024-160-CL. The video was obtained at $400$ frames per second, with a exposure time of $1.0$ ms per frame.
The filming was processed numerically to obtain the volume of the gas as a function of time. The pressure is registered
from the digital manometer with a computer through a data acquisition card National Instrument (DAQ-NI) USB-6008, that is also
used to synchronize the volume and pressure data. The ejection of water from the bottle lasted between
$0.3$ s to $0.7 $ s, depending on the initial pressure. The pressure-volume data of the expanding gas are adjusted with a
power law $PV^{\beta}=$constant in order to obtain an experimental value for $\beta$.

\begin{figure}[th]
\begin{center}
\includegraphics[scale=0.5, angle=0]{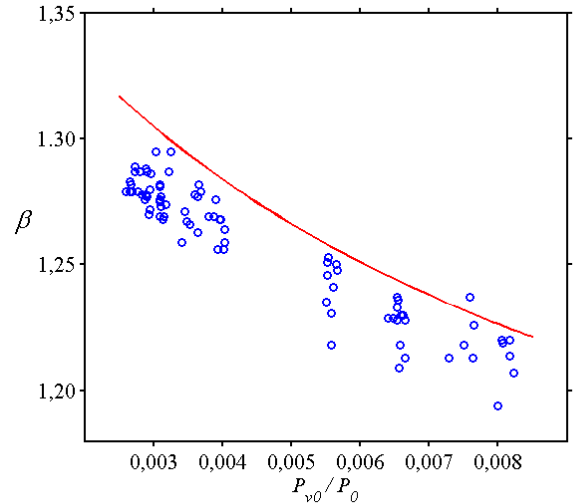}
\end{center}
\caption{The polytropic exponent as a function of the initial fraction between the partial pressure of vapor and the
initial wet air pressure. The circular points are the experimental values of $\beta$. The solid line is our theoretical
calculation given by Eq.(\ref{beta}).} \label{f2}
\end{figure}

Fig.~\ref{f2} shows the results for $90$ sets of experimental values of $\beta$ as a function of the initial temperature and the initial
pressure. Each set corresponds to different initial pressures with the same initial volume. The initial manometric
pressures used were: $2.2$, $3.0$, $3.8$, $5.0$, $6.0$, $6.5$, $7.0$, $7.5$ and $8.0$ bar (to obtain the absolute
pressure it is necessary to add the atmospheric pressure). For each initial pressure the experiment was repeated $10$ times
and all the data are incorporated in the same figure. The experimental values of the polytropic exponent $\beta$ are in
good agreement with the theoretical curve given by Eq.(\ref{beta}).

Our experiment cannot be reversed: we have no means to put the liquid water into the bottle quickly so
that the air is compressed and to be saturated with steam from the incoming water. However it is clear that this situation agrees, in the frame of our toy model of the tropical cyclone, with the fall of cold air from the tropopause to the ocean surface and its simultaneous compression.

\section{Polytropic processes}

A polytropic process is a quasistatic process carried out in such a way that the specific heat $c$ remains constant
during the entire process  \cite{Horedt,Drake,Chandra,Yong}. Therefore the heat exchange when the
temperature changes is
\begin{equation}
{dQ}=cN{dT},  \label{dqdt}
\end{equation}
where $Q$ is the heat absorbed by the system, $T$ is the temperature and $N$ is the number of moles. The processes in which
the pressure or the volume are kept constant are of course polytropic with specific heats $c_p$ or $c_{V} $ respectively.
In an adiabatic evolution there is no heat exchange between the system and the environment, therefore it is a polytropic
process with $c=0$. At the other extreme, the isothermal evolution is characterized by a constant temperature therefore
it may be thought as a polytropic process with infinite specific heat. Starting from the first law of thermodynamics and
using the ideal gas equation of state, it is easy to prove that the polytropic process satisfies
\begin{equation}
P {V}^\beta=constant,  \label{poly2}
\end{equation}
where $P$ is the pressure, $V$ the volume and
\begin{equation}
\beta=\frac{c_p-c}{c_{V}-c}.  \label{poly3}
\end{equation}
In particular for an adiabatic process $c=0$ and $\beta=\gamma\equiv{c_p}/c_{V}$; $\gamma=1.4$ for dry air. Using
Eq.(\ref{poly3}), the specific heat as a function of $\beta$ is
\begin{equation}
c=\frac{c_p-\beta c_{V}}{1-\beta}.  \label{poly4}
\end{equation}
The value of $c$ depends on the process and in principle it has no restriction \cite{Yong}; it
is seen that if $1<\beta<\gamma$ in Eq.(\ref{poly4}) then $c < 0$. 
In the experimental system treated in this paper we found $c<0$, this means that during the gas expansion heat is absorbed ($dQ>0$) yet
the gas temperature descends ($dT<0$) because the gas performs an amount of work larger than the absorbed heat. 

In ref.
\cite{romanelli} using the Clapeyron-Clausius equation  \cite{Reif} for the vapor-liquid phase transition the
following theoretical expression for the parameter $\beta$ was obtained
\begin{equation}
\beta =\frac{c_P/R+\left( L_{v}/R_{v}T_{0}\right) ^{2}(P_{v_{0}}/P_{0})}{%
c_{v}/R+\left( L_{v}/R_{v}T_{0}\right) ^{2}(P_{v_{0}}/P_{0})-\left( L_{v}/R_{v}T_{0}\right) (P_{v_{0}}/P_{0})}.
\label{beta}
\end{equation}%
where $R$ is the ideal gas constant, $L_v$ is the latent heat for the vapor-liquid phase transition, $R_v$ is the ideal gas constant
for the vapor, $T_{0}$ is the initial temperature, $P_{v_{0}}$ is the partial pressure of the vapor at the initial
temperature $T_{0}$, $P_{0}$ is the initial pressure of the atmosphere.
Analyzing Eq.(\ref{beta}) we conclude: (a) If $P_{v_{0}}\sim 0$ or $L_{v}\sim 0$ then $\beta \sim \gamma
=\frac{c_P}{c_{v}}$, this means that the process is adiabatic for dry air. (b) $\beta $ depends only on the
thermodynamical initial conditions of the gas.

\section{Maximum Potential Intensity and Carnot engine}

\begin{figure}[th]
\begin{center}
\includegraphics[scale=0.3, angle=0]{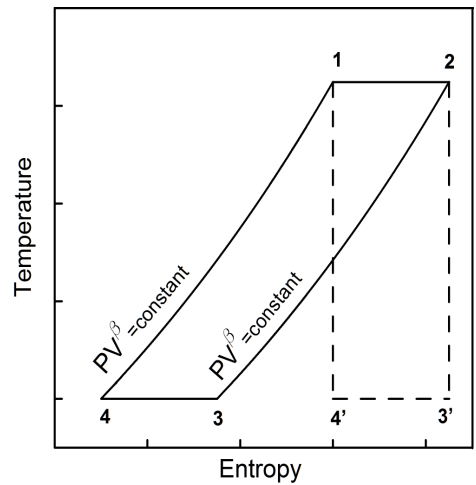}
\end{center}
\caption{In dashed line the usual Carnot cycle ($1,2,3^,,4^,$). In full line the modified Carnot cycle ($1,2,3,4$).
Process $2\rightarrow3$ represent the expansion of the wet air, that produces the water condensation associated with the
rain zone. Process $4\rightarrow1$ represent the compression of the cold dry air, that produces the water evaporation
associated with the dry zone. }\label{fa}
\end{figure}

The angular momentum of the tropical cyclone is only partially conserved due to friction with the sea surface. The
sea surface is a source of energy through water evaporation, however the friction also dissipates energy.
Using this information it is possible to obtain a theoretical upper bound for the wind speed $v_p$ \cite{Emanuel}.
This upper bound is called the "Maximum Potential Intensity", and is given by
\begin{equation}
v_p =\sqrt{\frac{C_k}{C_d}\epsilon({k_s}^*-k)} , \label{vp}
\end{equation}
where ${k_s}^*$ is the saturation enthalpy of the sea surface, $k$ is the enthalpy of the air boundary layer overlying the
surface, $C_k$ and $C_d$ are the exchange coefficients of enthalpy and momentum respectively and $\epsilon$ the
Carnot-engine efficiency given by
\begin{equation}
\epsilon=\frac{T_0-T_e}{T_0}
 ,\label{eps}
\end{equation}
where $T_0$ is the temperature of the sea surface and $T_e$ is the temperature of the tropopause. The Carnot engine goes
through a cycle consisting of four steps, see Fig.~\ref{fa}. (i) The inflowing air near the surface acquires energy
via the latent heat of evaporation of water at the constant temperature of the warm ocean surface. (ii) The
moist air ascends and expands adiabatically. (iii) The air gives off heat via
infrared radiation to space at the temperature of the cold tropopause. (iv) The air becomes compressed as it moves towards the
ground at the outer edge of the storm again adiabatically. In short, the first and third processes are
isothermal, while the second and fourth (when moist air rises or falls) are at constant entropy.

The adiabatic steps in the Carnot-cycle model of tropical cyclones are dubious because in storms
heat exchange is very important in these processes. In space these two
processes are developed simultaneously. There are two entangled air flows, one ascending and other descending
in a corkscrew shape. The water vapor condenses  in the up-flow, and an exchange of heat  is produced with the
cold down-flow. Accordingly, the main proposal of this work is to replace the Carnot cycle with a new modified cycle,
where the adiabatic processes are substituted by two polytropic processes with the same parameter $\beta$, see Fig.~\ref{fa}. In one of the polytropic
processes, the wet air is expanding and produces liquid water by condensation, the rain band. In the other polytropic process the
cold air falls, is compressed and evaporates water in its way, this is the cold dry band. 
Then, in the first process heat is delivered  and in the second the same amount of heat
is absorbed, between the same extreme temperatures and at the same rate $c$. Considered as a set,
the two polytropic processes do not exchange heat with the environment of
the tropical cyclone, the ocean and the tropopause.

This new proposed cycle works with only two heat sources: the warm ocean and the cold tropopause, then it
is easy to prove that the efficiency of the new cycle is the same of the usual Carnot cycle \cite{Sonntag}.
Therefore Eqs.(\ref{vp},\ref{eps}) remain valid with this new model,
whereby the maximum speed of the wind is maintained

\section{Discussion and conclusion}

We show a parallelism between the expansion and compression of the atmosphere in the secondary cycle of a tropical cyclone with the fast expansion and compression
of wet air in a bottle. In particular we suggest that the upward (downward) expansion of warm (cold), moist (dry) air
follows a polytropic process, $PV^\beta$= constant. We show both experimentally and analytically that $\beta$ depends on
the initial partial vapor pressure. The theoretical expression obtained could be used to evaluate the value of
$\beta$ for tropical cyclones.

We propose to change the adiabatic processes in the Carnot cycle model for the tropical cyclone by two polytropic
processes of expansion and compression of the air in the secondary cycle. These polytropic
processes can explain the existence of bands of rain and zones of dryness \cite{Houze}, spatially separated and
compatible with the heat exchange between the updraft and the downdraft. The rain band is associated with the polytropic
expansion of wet air and the dry band is associated with the polytropic compression of dry air. We show
that the efficiency of the new cycle is the same as that of the Carnot cycle and therefore the usual theoretical expression for the
maximum potential intensity is not modified.

Finally note that there remains the possibility to describe the updraft and the downdraft with different values of $\beta$ in which case the efficiency of the cycle would change, modifying the bound for the wind speed

A.R. and I.B. acknowledge stimulating discussions with V\'{\i}ctor Micenmacher, and the support
 of PEDECIBA and ANII, J.R. acknowledges the support of PEDECIBA through a scientific initiation scholarship.

\end{document}